%% file: main.tex
\documentclass[journal]{IEEEtran}
\IEEEoverridecommandlockouts
\usepackage{cite}
\usepackage[colorlinks=true,linkcolor=black,anchorcolor=black,citecolor=black,filecolor=black,menucolor=black,runcolor=black,urlcolor=magenta]{hyperref}

\usepackage{amsmath,amssymb,amsfonts,mathrsfs}
\usepackage{algorithmic}
\usepackage{graphicx}
\usepackage{textcomp}
\usepackage{xcolor}
\usepackage{booktabs}
\usepackage{subeqnarray}
\usepackage{multicol}
\usepackage[defaultcolor=red]{changes}
\def\BibTeX{{\rm B\kern-.05em{\sc i\kern-.025em b}\kern-.08em
    T\kern-.1667em\lower.7ex\hbox{E}\kern-.125emX}}

\ifCLASSOPTIONcompsoc
\usepackage[caption=false,font=normalsize,labelfont=sf,textfont=sf]{subfig}
\else
\usepackage[caption=false,font=footnotesize]{subfig}
\fi

\newcommand{\argmin}[1]{\underset{#1}{\operatorname{argmin}}\;}

\newcommand{\bs}[1]{\boldsymbol{#1}}

\newcommand{\stot}[1]{\mathrm{s2t}\left({#1}\right)}
\newcommand{\ttos}[1]{\mathrm{t2s}\left({#1}\right)}
\newcommand{\overbar}[1]{\mkern 1.5mu\overline{\mkern-1.5mu#1\mkern-1.5mu}\mkern 1.5mu}

\begin{document}
\bstctlcite{Settings}
\title{A Thru-free Multiline Calibration}

\author{%
	\IEEEauthorblockN{%
	Ziad~Hatab,~\IEEEmembership{Student~Member,~IEEE,}
	Michael~Ernst~Gadringer,~\IEEEmembership{Senior~Member,~IEEE,}\\ and~Wolfgang~Bösch,~\IEEEmembership{Fellow,~IEEE}
	}%
	\thanks{This work was supported by the Christian Doppler Research Association and by the Austrian Federal Ministry for Digital and Economic Affairs and the National Foundation for Research, Technology, and Development}
	\thanks{Ziad~Hatab, Michael~Ernst~Gadringer and Wolfgang~Bösch are with the Institute of Microwave and Photonic Engineering, Graz University of Technology, 8010 Graz, Austria, and also with the Christian Doppler Laboratory for Technology Guided Electronic Component Design and Characterization (TONI), 8010 Graz, Austria (e-mail: \{z.hatab, michael.gadringer, wbosch\}@tugraz.at).}
	\thanks{Software implementation and measurements are available online:\\ \url{https://github.com/ZiadHatab/thru-free-multiline-calibration}}
}%
\markboth{This work has been accepted for publication in the IEEE Transactions on Instrumentation and Measurement}{}
\maketitle

\begin{abstract}
This paper proposes a modification to the traditional multiline thru-reflect-line (TRL) or line-reflect-line (LRL) calibration method used for vector network analyzers (VNAs). Our proposed method eliminates the need for a thru (or line) standard by using an arbitrary transmissive two-port device in combination with an additional reflect standard. This combination of standards allows us to arbitrarily set the location of the calibration plane using physical artifacts. In contrast to the standard multiline TRL method, the suggested approach avoids a post-processing step to shift the calibration plane if a line standard is used. We demonstrate our proposed method with measurements on a printed circuit board (PCB) and compare it to the multiline TRL method with a perfectly defined thru.
\end{abstract}


\input{Sections/Section1}
\input{Sections/Section2}
\input{Sections/Section3}
\input{Sections/Section4}
\input{Sections/Section5}

\appendices
\input{Sections/AppendixA}
\input{Sections/AppendixB}

\section*{Acknowledgment}
The authors thank AT\&S, Leoben, Austria, for fabricating the PCB and ebsCENTER, Graz, Austria, for lending their equipment for the measurement.

\bibliographystyle{IEEEtran}
\bibliography{References/references.bib}

\end{document}

%% file: Sections/Section1.tex
\section{Introduction}
\label{sec:1}
\IEEEPARstart{T}{he} precision of measurements taken by a vector network analyzer (VNA) heavily relies on the calibration method's accuracy. Over the years, numerous improvements have been made to VNA calibration methods \cite{Rumiantsev2008}. Since its inception in 1979 \cite{Engen1979}, the thru-reflect-line (TRL) calibration method is still regarded as the most precise method for traceable VNA calibration. Although the TRL method is inherently bandlimited, an extension of the method called multiline TRL was proposed, which uses multiple line standards of varying lengths to expand the usable frequency range \cite{Marks1991}.

For both TRL and multiline TRL, a fully defined thru standard (a zero-length line) is required to determine the location of the calibration plane. However, in some applications, such a thru standard cannot be realized. For example, in on-wafer applications, the calibration plane should be at the tip of the probes \cite{Rumiantsev2020}. Undesirable effects can occur if the probes are placed too close to each other \cite{Orii2013,Phung2021}. In waveguide applications, the calibration plane is typically set at the adapter flanges. Although it is possible to create a thru standard by connecting the flanges directly, this results in a short length of the line standard at very high frequencies, which can be difficult to machine and handle \cite{Ridler2021,Ridler2019,Ridler2009}.

To avoid using a thru standard, a common solution is to define the calibration plane using a line standard of known length. Like the thru standard, this line must be fully specified. This method is called the line-reflect-line (LRL) method \cite{Hoer1987}. During the calibration process, the chosen line standard is treated as a thru standard, which places the calibration plane at the center of this line standard. The reference plane is then shifted to the desired location using the propagation constant extracted from the calibration procedure. The main challenge with this technique is the need for an accurate measurement of the propagation constant, which depends on knowledge of the exact length of the line standards. Additionally, the accuracy of the extracted propagation constant also depends on the choice of the length of the line standards. For example, a longer line may be useful in reducing uncertainty in the extracted propagation constant. However, a long line may be impractical due to physical limitations.

Another calibration method that does not require a thru standard is the short-open-load-reciprocal (SOLR) method \cite{Ferrero1992}. Unlike the LRL method, SOLR does not require a definition of a thru or line standard; instead, it uses any transmissive reciprocal device. With SOLR calibration, the location of the calibration plane is explicitly defined by the SOL standards at each port, which must be fully characterized. Therefore, the SOLR method's accuracy depends on the definition of the SOL standards.

Our proposed method eliminates the multiline calibration method's need for a thru standard. Instead, we use an arbitrary transmissive two-port device and an additional reflect standard to replace the thru standard. These standards physically define the location of the calibration plane. Although the suggested approach demands an additional reflect standard, all required standards are partially defined. This is in contrast to the multiline TRL (or LRL) method, where the thru (or line) standard is assumed to be perfectly defined.

The remainder of this article is organized as follows. Section \ref{sec:2} presents the application of the thru standard in multiline TRL calibration. In Section \ref{sec:3}, we derive the mathematical equations used to perform a thru-free multiline calibration. In Section \ref{sec:4} we experimentally compare our method with traditional multiline TRL calibration. Finally, we provide a summary in Section \ref{sec:5}. 


%% file: Sections/Section2.tex
\section{The Thru Standard in TRL calibration}
\label{sec:2}

\begin{figure}[th!]
	\centering
	\includegraphics[width=0.95\linewidth]{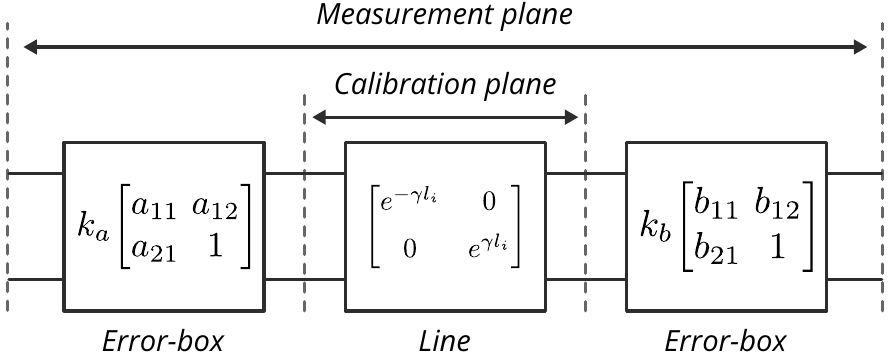}
	\caption{Two-port VNA error box model that illustrates the measurement of a line standard. All matrices are provided as T-parameters.}
	\label{fig:2.1}
\end{figure}

The error box model of a two-port VNA measuring a line standard is depicted in Fig.~\ref{fig:2.1}. The error box model can be simplified into seven terms as follows:
\begin{equation}
	\bs{M}_i = \underbrace{k_ak_b}_{k}\underbrace{\left[\begin{matrix}a_{11} & a_{12}\\[5pt]\
			a_{21} & 1\end{matrix}\right]}_{\bs{A}}\begin{bmatrix}
		e^{-\gamma l_i} & 0\\[5pt]
		0 & e^{\gamma l_i}
	\end{bmatrix} \underbrace{\left[\begin{matrix}b_{11} & b_{12}\\[5pt]
		b_{21} & 1\end{matrix}\right]}_{\bs{B}},
	\label{eq:2.1}
\end{equation}
where $\bs{A}$ and $\bs{B}$ are the one-port error boxes from each port, and $k$ is the 7th error term that describes the transmission between the two ports.

The first step in formulating TRL calibration is to set up the eigenvalue problem. This can be accomplished straightforwardly by taking measurements of two line standards with the same cross-section but different lengths (one of which can be a zero-length, i.e., thru). For example, the eigenvalue problem for the forward direction in terms of the matrix $\bs{A}$ is given by
\begin{equation}
    \bs{M}_i\bs{M}_j^{-1} = \bs{A}\begin{bmatrix}
        e^{-\gamma(l_i-l_j)} & 0\\[5pt]
        0 & e^{\gamma(l_i-l_j)}
    \end{bmatrix}\bs{A}^{-1}.
    \label{eq:2.2}
\end{equation}

This eigenvalue problem can also be applied in the reverse direction with respect to $\bs{B}$. Furthermore, a generalized weighted eigenvalue problem that combines multiple line standards at once can be derived, as discussed in \cite{Hatab2022}. In both the TRL and the multiline TRL calibration, the eigenvectors solve for the error boxes. Therefore, we can only solve for the error boxes in a normalized way, since eigenvectors are only unique up to a scalar factor. Specifically, we can obtain the following normalized error boxes from the eigenvectors.
\begin{equation}
	\widetilde{\bs{A}} = \begin{bmatrix}
		1 & a_{12} \\[5pt] 
		a_{21}/a_{11} & 1
	\end{bmatrix}, \qquad \widetilde{\bs{B}} = \begin{bmatrix}
	1 & b_{12}/b_{11} \\[5pt]
	b_{21} & 1
\end{bmatrix}.
	\label{eq:2.3}
\end{equation}

\begin{figure}[th!]
	\centering
	\includegraphics[width=0.95\linewidth]{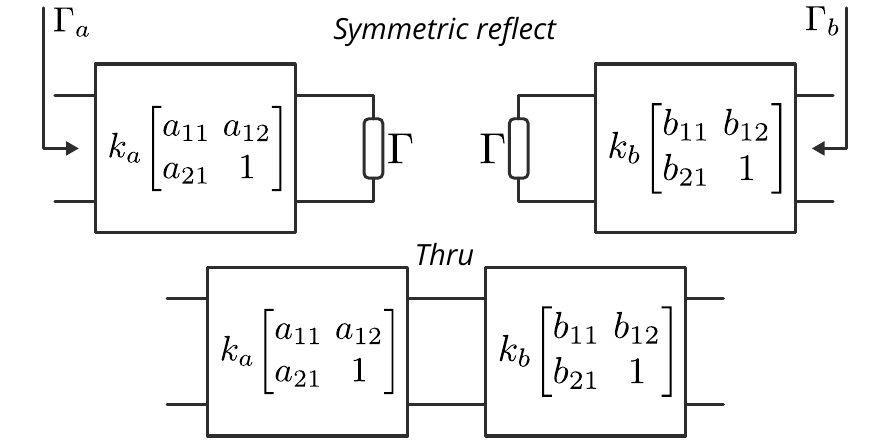}
	\caption{Two-port VNA error box model that illustrates the measurement of a symmetric reflect standard and a thru standard.}
	\label{fig:2.2}
\end{figure}

In order to recover all error terms of the VNA and denormalize the error boxes, we need to measure a thru standard and a symmetric reflect standard, as illustrated in Fig.~\ref{fig:2.2}. The thru standard is used to calculate the terms $k$ and $a_{11}b_{11}$, while the symmetric reflect standard is used to calculate the term $a_{11}/b_{11}$. By combining these terms with the normalized error terms, we can accurately recover all error terms. 

It is important to note that the normalized error terms obtained from the eigenvalue problem establish the reference impedance, which represents the characteristic impedance of the lines. On the other hand, the thru standard specifies the location of the reference plane, which is positioned at the center of the thru standard \cite{Marks1992}.

Using the measurement of the thru standard, we can calculate the terms $k$ and $a_{11}b_{11}$ directly by applying the normalized error boxes as follows:
\begin{equation}
	\widetilde{\bs{A}}^{-1}\bs{M}_\mathrm{thru}\widetilde{\bs{B}}^{-1} = \begin{bmatrix}
			ka_{11}b_{11} & 0 \\[5pt]
			0 & k
		\end{bmatrix}.
	\label{eq:2.4}
\end{equation}
where $a_{11}b_{11}$ is calculated by taking the ratio of the diagonal elements as $a_{11}b_{11} = ka_{11}b_{11}/k$.

Using the symmetrical reflect measurement, we can derive two equations, one for each port, that describe the input reflection coefficient. The equation for the left port (port $\bs{A}$) is as follows:
\begin{equation}
	\Gamma_a = \frac{a_{12}+a_{11}\Gamma}{1+a_{21}\Gamma}\quad\Longrightarrow\quad a_{11}\Gamma=\frac{\Gamma_a-a_{12}}{1-(a_{21}/a_{11})\Gamma_a},
	\label{eq:2.5}
\end{equation}
and from the right port (port $\bs{B}$) we have
\begin{equation}
	\Gamma_b = \frac{b_{11}\Gamma-b_{21}}{1-b_{12}\Gamma}\quad\Longrightarrow\quad b_{11}\Gamma=\frac{\Gamma_b+b_{21}}{1+(b_{12}/b_{11})\Gamma_b},
	\label{eq:2.6}
\end{equation}
where $\Gamma_a$ and $\Gamma_b$ are the raw measurements of the input reflection as seen from each port, and $\Gamma$ is the reflection coefficient of the symmetric reflect standard, which is not specified during calibration.

By combining both \eqref{eq:2.5} and \eqref{eq:2.6}, we can cancel the term $\Gamma$ and solve for $a_{11}/b_{11}$ as follows:
\begin{equation}
	\frac{a_{11}\Gamma}{b_{11}\Gamma}=\frac{a_{11}}{b_{11}} = \frac{\Gamma_a-a_{12}}{1-(a_{21}/a_{11})\Gamma_a}\frac{1+(b_{12}/b_{11})\Gamma_b}{\Gamma_b+b_{21}}.
	\label{eq:2.7}
\end{equation}

We can solve for $a_{11}$ and $b_{11}$ by using the values of $a_{11}b_{11}$ and $a_{11}/b_{11}$ as follows:
\begin{equation}
	a_{11} = \pm\sqrt{\frac{a_{11}}{b_{11}}a_{11}b_{11}}; \quad b_{11} = a_{11}\frac{b_{11}}{a_{11}}.
	\label{eq:2.8}
\end{equation}

To resolve the sign ambiguity, we select the answer closest to an estimate of $\Gamma$. We can apply the smallest Euclidean distance metric between the measured and estimated reflection coefficients to select the correct sign as summarized in \eqref{eq:2.9}.
\begin{equation}
	a_{11} = \argmin{a_{11}} \left\lbrace\left|\frac{\Gamma_a-a_{12}}{\pm a_{11}(1-(a_{21}/a_{11})\Gamma_a)} - \Gamma_\mathrm{est} \right| \right\rbrace.
	\label{eq:2.9}
\end{equation}

Finally, we denormalize the error boxes as follows:
\begin{subequations}
	\begin{align}
		\bs{A} =& \left[\begin{matrix}a_{11} & a_{12}\\[5pt]
			a_{21} & 1\end{matrix}\right] = \begin{bmatrix} 1 & a_{12}\\[5pt]
			a_{21}/a_{11} & 1\end{bmatrix}\begin{bmatrix} a_{11} & 0\\[5pt]
			0 & 1\end{bmatrix}\\[5pt]
		\bs{B} =& \left[\begin{matrix}b_{11} & b_{12}\\[5pt]
			b_{21} & 1\end{matrix}\right] = \begin{bmatrix} b_{11} & 0\\[5pt]
			0 & 1\end{bmatrix}\begin{bmatrix} 1 & b_{12}/b_{11}\\[5pt]
			b_{21} & 1\end{bmatrix}.
	\end{align}
	\label{eq:2.10}
\end{subequations}

In summary, if we can compute the terms $k$ and $a_{11}b_{11}$ without relying on the availability of a thru standard, we have achieved our goal.


%% file: Sections/Section3.tex
\section{Derivation of Thru-free Calibration}
\label{sec:3}

\begin{figure}[th!]
	\centering
	\includegraphics[width=0.95\linewidth]{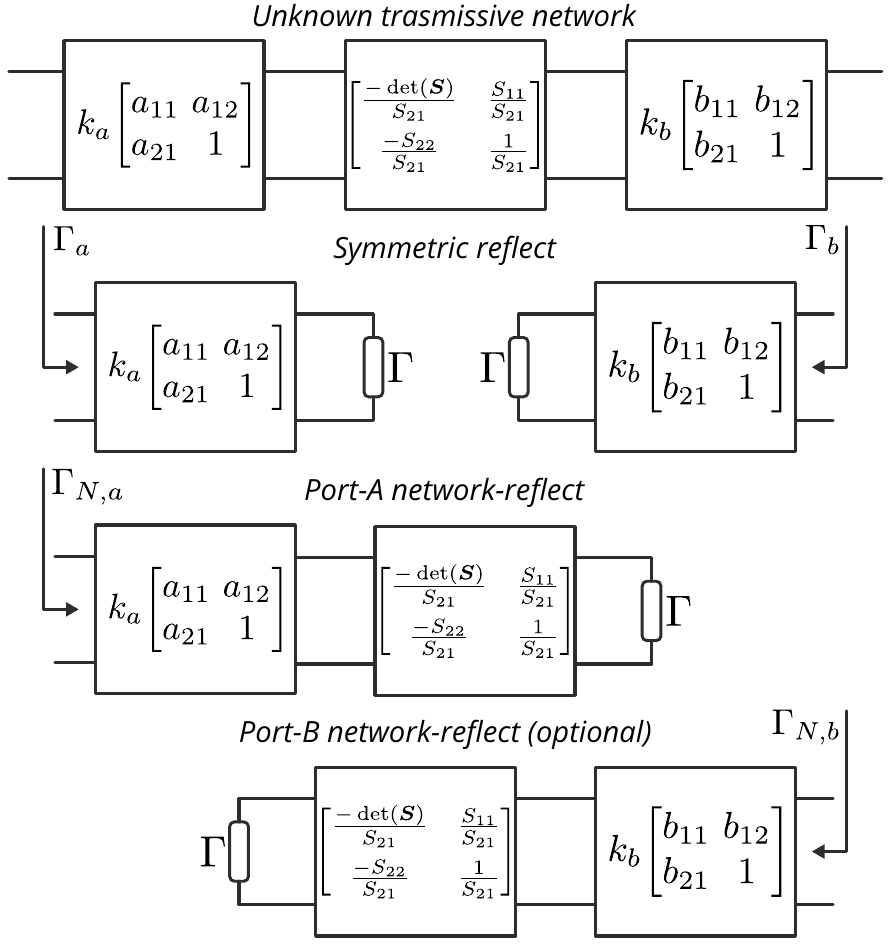}
	\caption{Error box model of the required standards for the denormalization of the error terms in the thru-free calibration method.}
	\label{fig:3.1}
\end{figure}

Instead of explicitly defining a thru standard, we combine a reflect standard with an unspecified two-port network standard. We assume that the eigenvalue problem from the various line standards has already been solved and that the normalized error terms have been derived. To perform the denormalization and determine the error terms $a_{11}$ and $b_{11}$, we use the standards shown in Fig.~\ref{fig:3.1}.

To derive $a_{11}b_{11}$, it is not necessary that the unknown network be reciprocal. Any transmissive network (i.e., $|S_{12}|, |S_{21}| > 0$) will suffice. By applying the normalized error boxes to the network's measurement, we obtain the following expression:
\begin{equation}
    \widetilde{\bs{A}}^{-1}\bs{M}_\mathrm{net}\widetilde{\bs{B}}^{-1} = k\begin{bmatrix} a_{11} & 0\\[5pt]
		0 & 1\end{bmatrix}\begin{bmatrix}
		\frac{-\det\left(\bs{S}\right)}{S_{21}} & \frac{S_{11}}{S_{21}}\\[5pt]
		\frac{-S_{22}}{S_{21}} & \frac{1}{S_{21}}
	\end{bmatrix}\begin{bmatrix} b_{11} & 0\\[5pt]
	0 & 1\end{bmatrix},
	\label{eq:3.1}
\end{equation}
where $\det\left(\bs{S}\right) = S_{11}S_{22} - S_{21}S_{12}$. Converting back to S-parameters yields the following result:
\begin{equation}
    \ttos{\widetilde{\bs{A}}^{-1}\bs{M}_\mathrm{net}\widetilde{\bs{B}}^{-1}} = \begin{bmatrix}
		a_{11}S_{11} & a_{11}b_{11}S_{12}k\\[5pt]
		S_{21}/k & b_{11}S_{22}
	\end{bmatrix}.
	\label{eq:3.2}
\end{equation}

From the symmetric reflect measurement, we can derive two equations similar to the TRL calibration as presented in \eqref{eq:2.5} and \eqref{eq:2.6}:
\begin{equation}
	a_{11}\Gamma=\frac{\Gamma_a-a_{12}}{1-(a_{21}/a_{11})\Gamma_a}, \quad b_{11}\Gamma=\frac{\Gamma_b+b_{21}}{1+(b_{12}/b_{11})\Gamma_b}.
	\label{eq:3.3}
\end{equation}

Finally, we use the last standard, which is the network-reflect standard. For the left configuration (i.e., port $\bs{A}$), we can derive the input reflection coefficient in a similar way to the previous case, by recognizing that the reflect standard is cascaded with the unknown network. This is given as follows:
\begin{equation}
	a_{11}\frac{\Gamma S_{11} S_{22} - \Gamma S_{12} S_{21} - S_{11}}{\Gamma S_{22} - 1}=\frac{\Gamma_{N,a}-a_{12}}{1-(a_{21}/a_{11})\Gamma_{N,a}}.
	\label{eq:3.4}
\end{equation}

A similar equation can be derived if we consider the measurement from the right port (i.e., port $\bs{B}$), which is given as follows:
\begin{equation}
	b_{11}\frac{\Gamma S_{11}S_{22} - \Gamma S_{12}S_{21} - S_{22}}{\Gamma S_{11} - 1} = \frac{\Gamma_{N,b}+b_{21}}{1+(b_{12}/b_{11})\Gamma_{N,b}}.
	\label{eq:3.5}
\end{equation}

From \eqref{eq:3.2}--\eqref{eq:3.5}, we can summarize the following seven equations relating the model and measurement:
\begin{subequations}
	\begin{align}
		m_1 &= a_{11}\Gamma,\label{eq:3.6a}\\
		m_2 &= a_{11}S_{11},\label{eq:3.6b}\\
		m_3 &= b_{11}\Gamma,\label{eq:3.6c}\\
		m_4 &= b_{11}S_{22},\label{eq:3.6d}\\
		m_5 &= a_{11}b_{11}S_{21}S_{12},\label{eq:3.6e}\\
		m_6 &= \frac{a_{11} \left(\Gamma S_{11} S_{22} - \Gamma S_{12} S_{21} - S_{11}\right)}{\Gamma S_{22} - 1},\label{eq:3.6f}\\
		m_7 &= \frac{b_{11} \left(\Gamma S_{11}S_{22} - \Gamma S_{12}S_{21} - S_{22}\right)}{\Gamma S_{11} - 1}.\label{eq:3.6g}
	\end{align}
	\label{eq:3.6}
\end{subequations}

The value of $m_5$ in \eqref{eq:3.6e} was calculated by multiplying the off-diagonal elements of the S-parameters in \eqref{eq:3.2}.

We begin the derivation of $a_{11}b_{11}$ with the measurement of $m_6$ from \eqref{eq:3.6f}. First, we distribute $a_{11}$ over the numerator,
\begin{equation}
	m_6 = \frac{a_{11}\Gamma S_{11}S_{22} - a_{11}\Gamma S_{12}S_{21} - S_{11}a_{11}}{\Gamma S_{22} - 1}.
	\label{eq:3.7}
\end{equation}

Then, we substitute $m_1 = a_{11}\Gamma$ and $m_2 = a_{11}S_{11}$, which gives us
\begin{equation}
	m_6 = \frac{m_1 S_{11}S_{22} - m_1 S_{12}S_{21} - m_2}{\Gamma S_{22} - 1}.
	\label{eq:3.8}
\end{equation}

Subsequently, we multiply both the numerator and the denominator by $a_{11}b_{11}$. This gives us the following expression:
\begin{equation}
	m_6 = \frac{m_1a_{11}b_{11}S_{11}S_{22} - m_1a_{11}b_{11}S_{12}S_{21} - m_2a_{11}b_{11}}{a_{11}b_{11}\Gamma S_{22} - a_{11}b_{11}}.
	\label{eq:3.9}
\end{equation}

We simplify the above expression by substituting the corresponding values of $m_1$, $m_2$, $m_3$, $m_4$, and $m_5$. This results in the following expression in terms of $a_{11}b_{11}$:
\begin{equation}
	m_6 = \frac{m_1m_2m_4 - m_1m_5 - m_2a_{11}b_{11}}{m_1m_4 - a_{11}b_{11}}.
	\label{eq:3.10}
\end{equation}

Lastly, we rearrange the above expression and solve for $a_{11}b_{11}$ as follows:
\begin{equation}
	a_{11}b_{11} = \frac{m_1m_2m_4-m_1m_5-m_6m_1m_4}{m_2-m_6}.
	\label{eq:3.11}
\end{equation}

The above expression for $a_{11}b_{11}$ can be further simplified as follows:
\begin{equation}
	a_{11}b_{11} = m_1m_4 - \frac{m_1m_5}{m_2-m_6}.
	\label{eq:3.12}
\end{equation}

It is worth noting that in \eqref{eq:3.12}, we did not need the measurement $m_7$ from \eqref{eq:3.6g}. However, the same process can be done for $m_7$ without the need for $m_6$. Basically, in the above result, we swap $m_6 \leftrightarrow m_7$, $m_1 \leftrightarrow m_3$, and $m_2 \leftrightarrow m_4$. This results in the alternative solution for $a_{11}b_{11}$ as follows:
\begin{equation}
	a_{11}b_{11} = m_3m_2 - \frac{m_3m_5}{m_4-m_7}.
	\label{eq:3.13}
\end{equation}

If both $m_6$ and $m_7$ are available, we can establish an average measurement for $a_{11}b_{11}$ or compare the two results of $a_{11}b_{11}$ for a calibration consistency check. Once $a_{11}b_{11}$ has been solved, we can use equations \eqref{eq:2.7}--\eqref{eq:2.9} to solve for $a_{11}$ and $b_{11}$, and then denormalize the error boxes using \eqref{eq:2.10}.

To complete the calibration, we only need to solve for the transmission error term $k$. We can use the same method as in SOLR calibration \cite{Ferrero1992} by calculating $k$ through the determinate of the single-port corrected measurement of a two-port reciprocal device, i.e., $S_{21}=S_{12}$. For a reciprocal network, like the line standards, the calibrated measurement by the single-port error boxes is given by:
\begin{equation}
	\bs{A}^{-1}\bs{M}_\mathrm{recip}\bs{B}^{-1} = \frac{k}{S_{21}}\begin{bmatrix}
		S_{21}^2 - S_{11}S_{22} & S_{11}\\[5pt]
		-S_{22} & 1
	\end{bmatrix}.
	\label{eq:3.14}
\end{equation}

By taking the determinant from both sides, we obtain
\begin{equation}
	\det\left(\bs{A}^{-1}\bs{M}_\mathrm{recip}\bs{B}^{-1}\right) = k^2.
	\label{eq:3.15}
\end{equation}

Hence, $k$ is solved as follows:
\begin{equation}
	k = \pm\sqrt{\det\left(\bs{A}^{-1}\bs{M}_\mathrm{recip}\bs{B}^{-1}\right)}
	\label{eq:3.16}
\end{equation}

To determine the appropriate sign, we choose the answer closest to a known estimate of the reciprocal network. This estimate could be based on the line standard through the estimated value of the propagation constant or material properties. Furthermore, since all line standards are reciprocal, we can compute $k^2$ from all of them and determine an average value.

In Table \ref{tab:3.1}, we present a summary comparison of the definition of standards in the multiline TRL calibration and the thru-free calibration.
\begin{table}[ht!]
	\centering
	\caption{Comparison of standard definition in multiline TRL and thru-free calibrations.}
	\label{tab:3.1}
	\begin{tabular}{@{$\ $}ccccc@{$\ $}}
		\toprule
		&
		\begin{tabular}[c]{@{}c@{}}Thru \\ (or line)\end{tabular} &
		Lines &
		\begin{tabular}[c]{@{}c@{}}Symmetric \\ Reflect\end{tabular} &
		Network \\ \midrule
		\begin{tabular}[c]{@{}c@{}}Multiline \\ TRL\end{tabular} &
		\begin{tabular}[c]{@{}c@{}}All $S_{ij}$ \\ must be \\ specified\end{tabular} &
		\begin{tabular}[c]{@{}c@{}}$S_{11}=0$\\ $S_{22}=0$\\ $S_{21}=S_{12}$\end{tabular} &
		$S_{11}=S_{22}$ &
		Not used \\ \midrule
		Thru-free &
		Not used &
		\begin{tabular}[c]{@{}c@{}}$S_{11}=0$\\ $S_{22}=0$\\ $S_{21}=S_{12}$\end{tabular} &
		$S_{11}=S_{22}$ &
		\begin{tabular}[c]{@{}c@{}}Arbitrary, \\ as long as\\ $|S_{21}|,|S_{12}|>0$\end{tabular} \\ \bottomrule
	\end{tabular}
\end{table}

\begin{figure}[th!]
	\centering
	\includegraphics[width=1\linewidth]{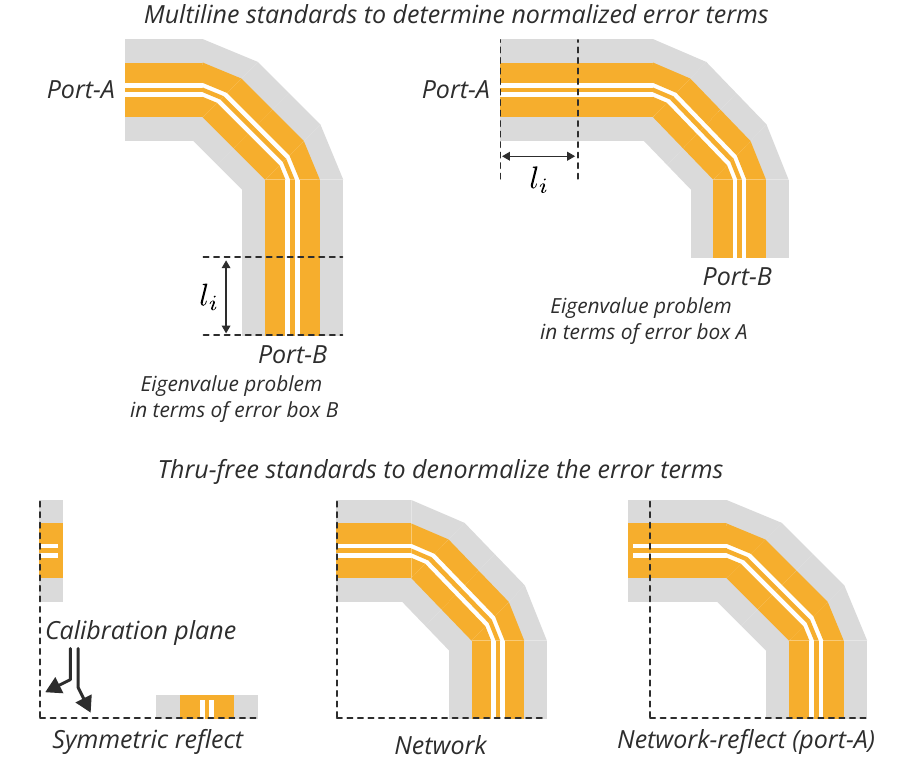}
	\caption{Illustration example of thru-free multiline calibration of CPW standards with a 90\textdegree\ bend.}
	\label{fig:3.2}
\end{figure}

A potential use for the proposed thru-free method is calibrating at various bend angles, which may be necessary for on-wafer applications. To accomplish this, we must first calculate the normalized error boxes as defined in \eqref{eq:2.3}. Since there is a bend, we need to perform the eigendecomposition on two different sets of lines. This process is achieved in two steps. First, we sweep the line at port-A and estimate the normalized error box for port-A. Second, we sweep the line at port-B to determine the normalized error box for port-B. This process is illustrated in Fig.~\ref{fig:3.2}. Afterwards, we apply the proposed method by measuring a network standard that directly connects to the bending element, along with a symmetric reflect standard at the desired calibration plane and an additional network-reflect standard at either port. A full example of such standards using co-planar waveguide (CPW) is provided in Fig.~\ref{fig:3.2}.


%% file: Sections/Section4.tex
\section{Experiment}
\label{sec:4}
 
\subsection{Measurement setup}
In this experiment, we fabricated a set of multiline standards as microstrip lines on a printed circuit board (PCB). The PCB consists of four copper layers, with the top two layers used for the fabricated microstrip lines. The substrate material is Panasonic Megtron 7, with a specified dielectric constant of 3.4 and a tangent loss of 0.002. The multiline TRL kit includes multiple microstrip lines with lengths of $\{0, 0.5, 1, 1.5, 2, 3, 5, 6.5\}\,\mathrm{mm}$, and a reflect standard implemented as a short using microvias. The microstrip lines' probing pads are implemented using a low-return loss design of ground-signal-ground (GSG) pads, as discussed in \cite{Hatab2022b}. The microstrip lines have a width of $0.107\,\mathrm{mm}$ and a substrate thickness of $0.05\,\mathrm{mm}$, corresponding to an average characteristic impedance of $50\,\Omega$.

We use the same line and reflect standards for the thru-free kit as in the multiline TRL kit. Additionally, we use a network standard implemented as a $1\,\mathrm{mm}$ line and a network-reflect standard implemented as an offset short, which is implemented using the same microvia, offsetted by $1\,\mathrm{mm}$. The network-reflect standard is implemented for both ports to demonstrate that the usage of either port will result in the same solution.

\begin{figure}[th!]
	\centering
	\includegraphics[width=0.95\linewidth]{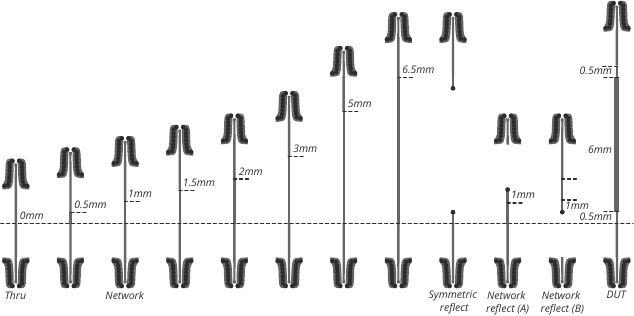}
	\caption{Schematic illustration of the measured structures.}
	\label{fig:4.1x}
\end{figure}

In addition to the calibration standards, we included a device under test (DUT) for comparison purposes. The DUT is implemented as a stepped-impedance line with a length of $6\,\mathrm{mm}$ and a width of $0.22\,\mathrm{mm}$, corresponding to an average characteristic impedance of $30\,\Omega$. The DUT is placed at an offset of $0.5\,\mathrm{mm}$. A schematic of all measured structures is shown in Fig.~\ref{fig:4.1x}.

\begin{figure}[th!]
	\centering
	\includegraphics[width=0.95\linewidth]{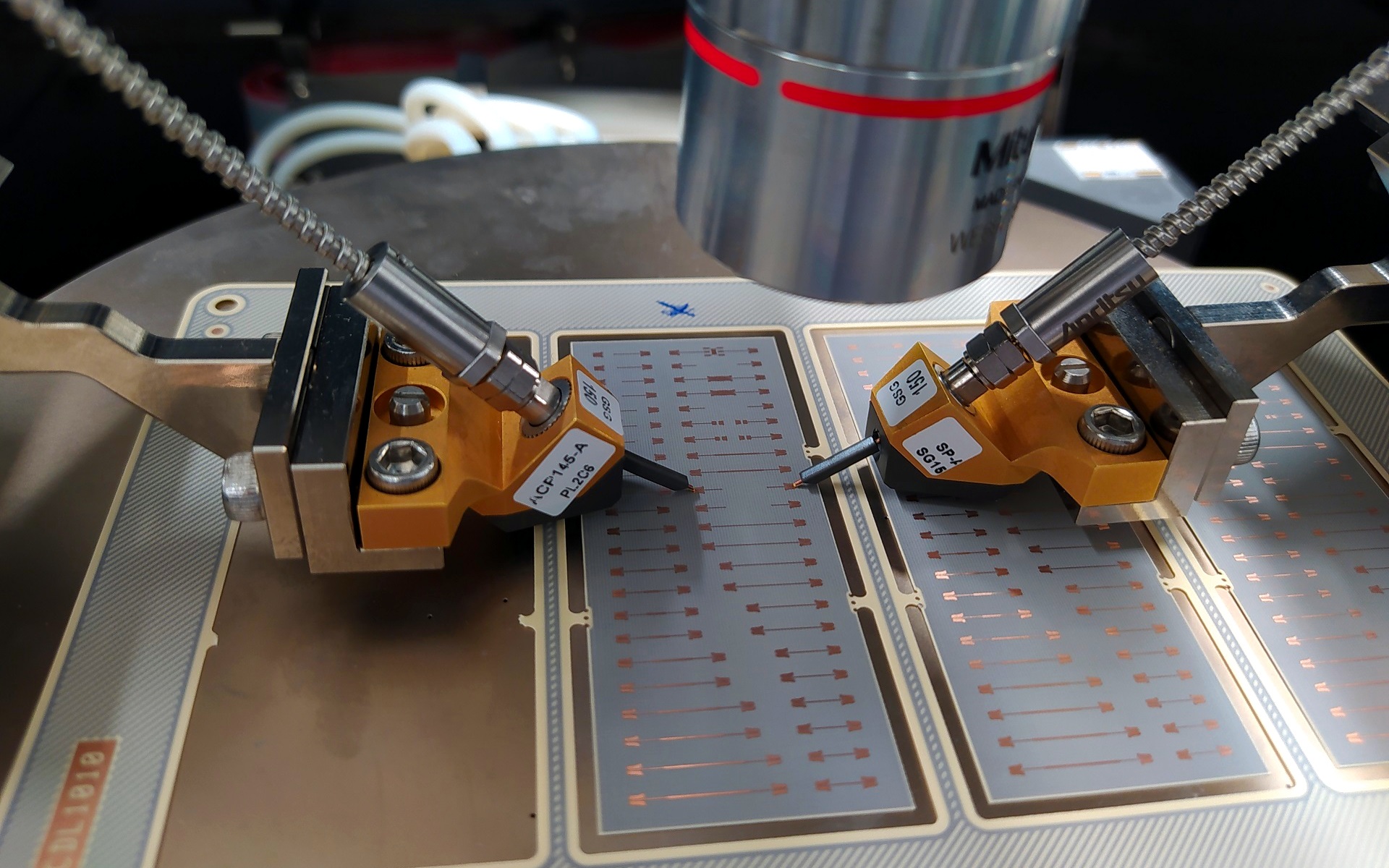}
	\caption{Measurement setup depicting the ACP probes and the PCB carrying the calibration standards and DUT.}
	\label{fig:4.1}
\end{figure}

The instrumentation setup consists of an Anritsu VectorStar VNA with millimeter-wave extensions to support frequencies up to $150\,\mathrm{GHz}$. The probes used are ACP probes from FormFactor with a GSG-pitch of $150\,\mu\mathrm{m}$. The measurement was performed on the SUMMIT200 probe station. A photograph of the measurement setup is shown in Fig.~\ref{fig:4.1}.

\subsection{Results and discussion}
The raw S-parameter measurements of the calibration standards were collected over multiple frequency sweeps. For each standard, 25 frequency sweeps were collected at an IF-bandwidth of $100\,\mathrm{Hz}$ and a source power of $-10\,\mathrm{dBm}$. Each frequency sweep covers the range $1-150\,\mathrm{GHz}$ with 299 frequency points. The collected data was processed in Python with help of the package \textit{scikit-rf} \cite{Arsenovic2022}, and the multiline TRL algorithm from \cite{Hatab2022} was used. We also applied the same eigenvalue formulation from \cite{Hatab2022} for the thru-free multiline calibration. Both methods result in the same normalized error terms, as a thru definition is not required in the formulation of the eigenvalue problem. We denormalized the error terms for the multiline TRL calibration using the reflect standard (short) and the thru standard ($0\,\mathrm{mm}$ line) to define the location of the calibration plane at the center of the thru standard. Thereafter, we used the same reflect standard (short) for the thru-free calibration, in addition to the network standard implemented as a $1\,\mathrm{mm}$ line and the network-reflect standard implemented as an offset short, with the offset being identical to the network standard ($1\,\mathrm{mm}$ line). Furthermore, since we collected multiple sweeps for each standard, we computed the covariance matrix due to instrument noise and linearly propagated its uncertainty through both calibrations using the technique discussed in \cite{Hatab2022a,Hatab2023}.

\begin{figure}[th!]
	\centering
	\includegraphics[width=1\linewidth]{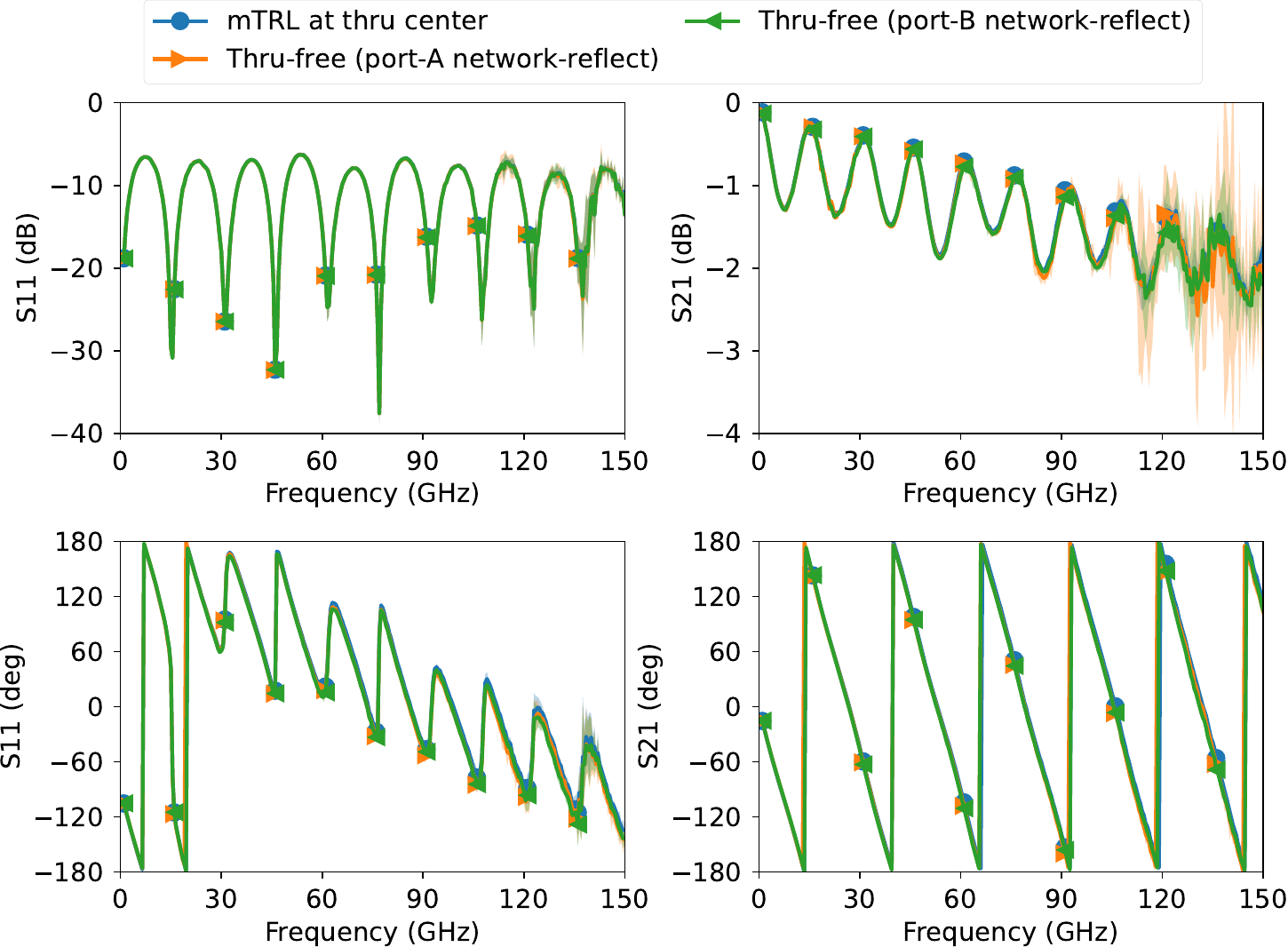}
	\caption{The calibrated measurement of the $6\,\mathrm{mm}$ long $30\,\Omega$ stepped-impedance line. The calibrated measurement of $S_{22}$ and $S_{12}$ are not shown, as they behave similarly to $S_{11}$ and $S_{21}$. The uncertainty bounds correspond to a 95\,\% coverage of a Gaussian distribution due to noise from the VNA propagated linearly through the calibrations.}
	\label{fig:4.2}
\end{figure}

In Fig.~\ref{fig:4.2}, we show the S-parameters of the calibrated DUT ($6\,\mathrm{mm}$ long $30\,\Omega$ stepped-impedance line) using both calibration methods. For the thru-free method, we investigated both cases when using the network-reflect standard from either port. Generally, both the multiline TRL and the thru-free calibration methods show overlapping agreement. However, when we look at the uncertainty bounds, we see that for the calibrated $S_{11}$, we obtain similar uncertainty bounds for both calibration methods, whereas for the calibrated $S_{21}$ measurement, we see that the uncertainty in the magnitude is slightly higher for the thru-free method at frequencies above $110\,\mathrm{GHz}$. More notability, the uncertainty of the thru-free method is much higher when using the network-reflect standard at port $\bs{A}$.

In Table~\ref{tab:4.1}, we summarize the mean absolute error across all frequencies of the magnitude and phase of the calibrated DUT by the thru-free method with respect to multiline TRL. We can observe that the error of the thru-free method when using the network-reflect at port-A is slightly higher than when using the network-reflect at port-B. The equations for computing the error metric are given as follows:
\begin{subequations}
	\begin{align}
		\overbar{\Delta|S_{ij}|_\mathrm{dB}} =& \frac{1}{M_f}\sum_{f_i}\left||S_{ij}^\mathrm{free}(f_i)|_\mathrm{dB}-|S_{ij}^\mathrm{mTRL}(f_i)|_\mathrm{dB}\right|\\
		\overbar{\Delta\arg(S_{ij})} =& \frac{1}{M_f}\sum_{f_i}\left|\arg(S_{ij}^\mathrm{free}(f_i))-\arg(S_{ij}^\mathrm{mTRL}(f_i))\right|
	\end{align}
	\label{eq:4.1}
\end{subequations}
where $M_f$ is the total number of frequency points.
\begin{table}[ht!]
	\centering
	\caption{Mean absolute error of magnitude and phase of the thru-free method with respect to the multiline TRL based on \eqref{eq:4.1}.}
	\label{tab:4.1}
	\begin{tabular}{@{$\quad$}ccccc@{$\quad$}}
		\toprule
		&
		$\overbar{\Delta|S_{11}|_\mathrm{dB}}$ &
		$\overbar{\Delta\arg(S_{11})}$ &
		$\overbar{\Delta|S_{21}|_\mathrm{dB}}$ &
		$\overbar{\Delta\arg(S_{21})}$ \\ \midrule
		\begin{tabular}[c]{@{}c@{}}Thru-free \\ (port-A)\end{tabular} &
		0.062 &
		5.187\textdegree &
		0.061 &
		5.098\textdegree \\ \midrule
		\begin{tabular}[c]{@{}c@{}}Thru-free \\ (port-B)\end{tabular} &
		0.059 &
		5.090\textdegree &
		0.059 &
		5.003\textdegree \\ \bottomrule
	\end{tabular}
\end{table}

The noise impact on the thru-free calibration becomes no- noticeable after $110\,\mathrm{GHz}$, but the calibration algorithm does not cause this.  Instead, it is attributed to the VNA itself, specifically its poor performance at port 1 (i.e., port $\bs{A}$). Measurements taken at this port are always noisier compared to the opposite port, which explains why the uncertainty bounds are much higher when using the network-reflect at port $\bs{A}$ than when using network-reflect standard at port $\bs{B}$. Appendix~\ref{anx:A} provides a more detailed analysis of the noise imbalance between the ports of the Anritsu ME7838D VNA.

\begin{figure}[th!]
	\centering
	\includegraphics[width=1\linewidth]{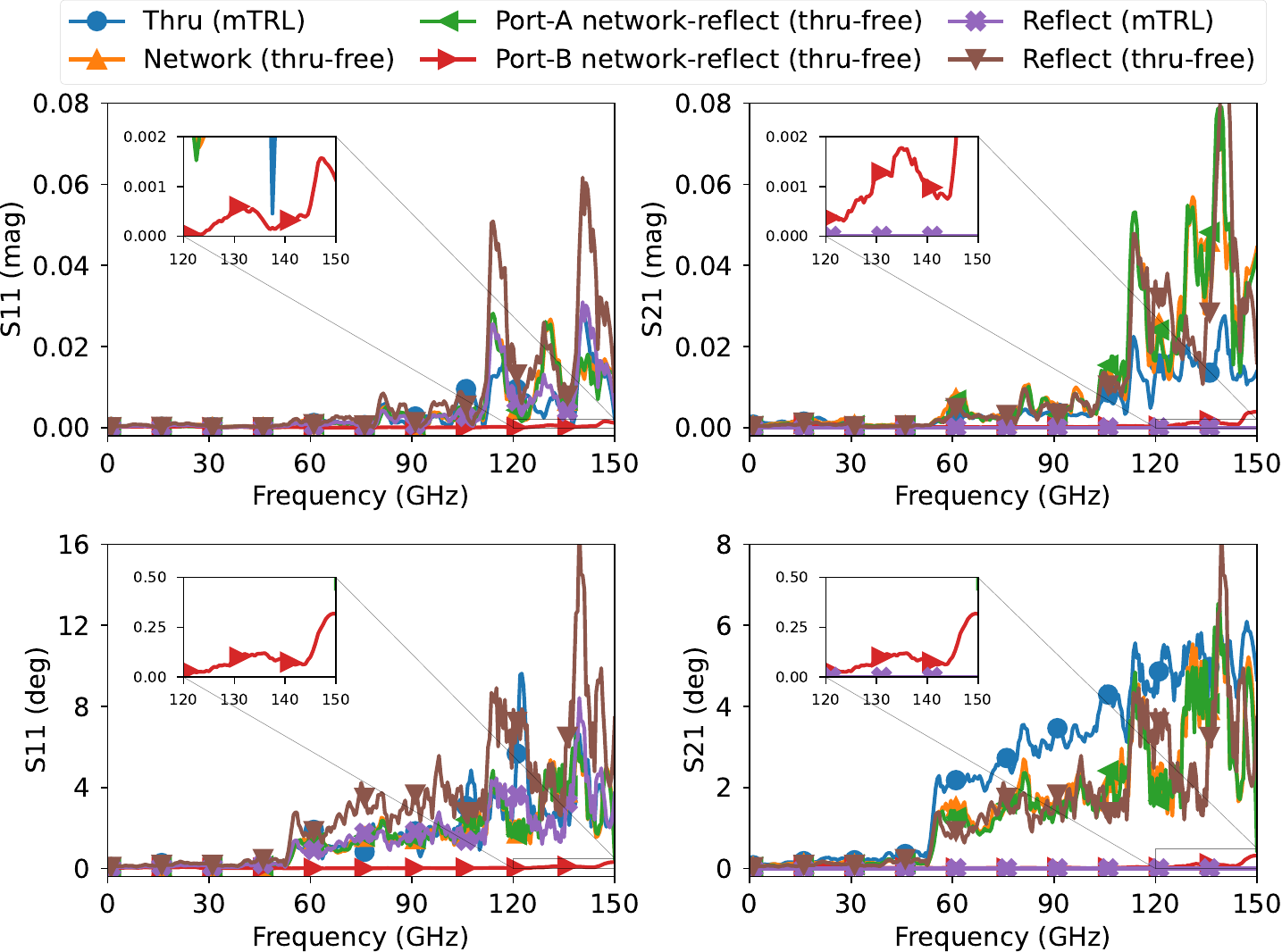}
	\caption{Uncertainty budget of the calibrated stepped-impedance line due to the calibration standards. The uncertainty is represented as 95\,\% coverage of a Gaussian distribution. The traces have been smoothed for readability using a Savitzky-Golay filter \cite{Savitzky1964} with a window size of 9 and a polynomial order of 2.}
	\label{fig:4.3}
\end{figure}

While the noise sensitivity between different ports is directly related to the VNA, it is still important to analyze the uncertainty contribution from each calibration standard due to VNA noise to the calibrated DUT. To do this, we consider the uncertainty budget due to each standard in the calibrated DUT. Both calibration methods use the same line standards in the exact same way in formulating the eigenvalue problem, therefore, these standards are not included in the budget analysis. Instead, we consider the thru and reflect standards for the multiline TRL calibration and the reflect, network, and network-reflect standards for the thru-free method. In Fig.~\ref{fig:4.3}, we show the uncertainty contribution from these standards to the calibrated S-parameters of the DUT. For the magnitude response, we have plotted the uncertainties in linear scale, as it is easier to interpret than in the dB scale. Additionally, for clarity, we included in Table~\ref{tab:4.2} the uncertainty budget due to the standards at the frequency $110\,\mathrm{GHz}$.

Regarding the uncertainties in $S_{11}$, all standards exhibit similar contributions in terms of magnitude and phase, except for the network-reflect standard at port $\bs{B}$, which is a single-port measurement that inherently has less noise than the other port. It should be noted that the reflect standard for both multiline TRL and thru-free method is a two-port measurement.  Hence the high noise from port $\bs{A}$ is present. As for the uncertainty contribution in $S_{21}$, we observe that all calibration standards contribute to the uncertainty for the thru-free method. In contrast, for multiline TRL calibration, the reflect standard has no impact at all. This behavior may seem counterintuitive since the reflect standard is part of the calibration. However, this result is not surprising since the reflect standard contributes to deriving the ratio error term $a_{11}/b_{11}$, which, in turn, allows the separation of the error terms $a_{11}$ and $b_{11}$. We can demonstrate that the calibrated $S_{21}$ can be entirely calculated without the requirement of the reflect standard. This is because only the normalized error terms $\{a_{12}, a_{21}/a_{11}, b_{21}, b_{12}/b_{11}\}$, the combined error term $a_{11}b_{11}$, and the transmission error term $k$ are needed to describe the calibrated $S_{21}$ response. A derivation of this relationship is presented in Appendix~\ref{anx:B}.

\begin{table}[ht!]
	\centering
	\caption{Uncertainty budget due to each standard in the calibrated DUT at $110\,\mathrm{GHz}$ as provided in Fig.~\ref{fig:4.3}.}
	\label{tab:4.2}
	\begin{tabular}{@{$\quad$}ccccc@{$\quad$}}
		\toprule
		&
		\begin{tabular}[c]{@{}c@{}}$\mathrm{unc}(S_{11})$\\ (mag)\end{tabular} &
		\begin{tabular}[c]{@{}c@{}}$\mathrm{unc}(S_{11})$\\ (deg)\end{tabular} &
		\begin{tabular}[c]{@{}c@{}}$\mathrm{unc}(S_{21})$\\ (mag)\end{tabular} &
		\begin{tabular}[c]{@{}c@{}}$\mathrm{unc}(S_{21})$\\ (deg)\end{tabular} \\ \midrule
		Thru                                                                & 0.0032 & 0.912 & 0.0069 & 3.8905 \\\midrule
		\begin{tabular}[c]{@{}c@{}}Reflect\\ (mTRL)\end{tabular}            & 0.0022 & 1.7218 & 0.0    & 0.0    \\\midrule
		Network                                                             & 0.0038 & 2.7492 & 0.0141 & 2.9622 \\\midrule
		\begin{tabular}[c]{@{}c@{}}Reflect\\ (Thru-free)\end{tabular}       & 0.0043 & 3.4455 & 0.0077 & 1.7239 \\\midrule
		\begin{tabular}[c]{@{}c@{}}Network-reflect \\ (Port-A)\end{tabular} & 0.0044 & 2.5652 & 0.0154 & 2.5652 \\\midrule
		\begin{tabular}[c]{@{}c@{}}Network-reflect \\ (Port-B)\end{tabular} & 0.0001 & 0.0236 & 0.0004 & 0.0236 \\ \bottomrule
	\end{tabular}
\end{table}

\begin{figure}[th!]
	\centering
	\includegraphics[width=1\linewidth]{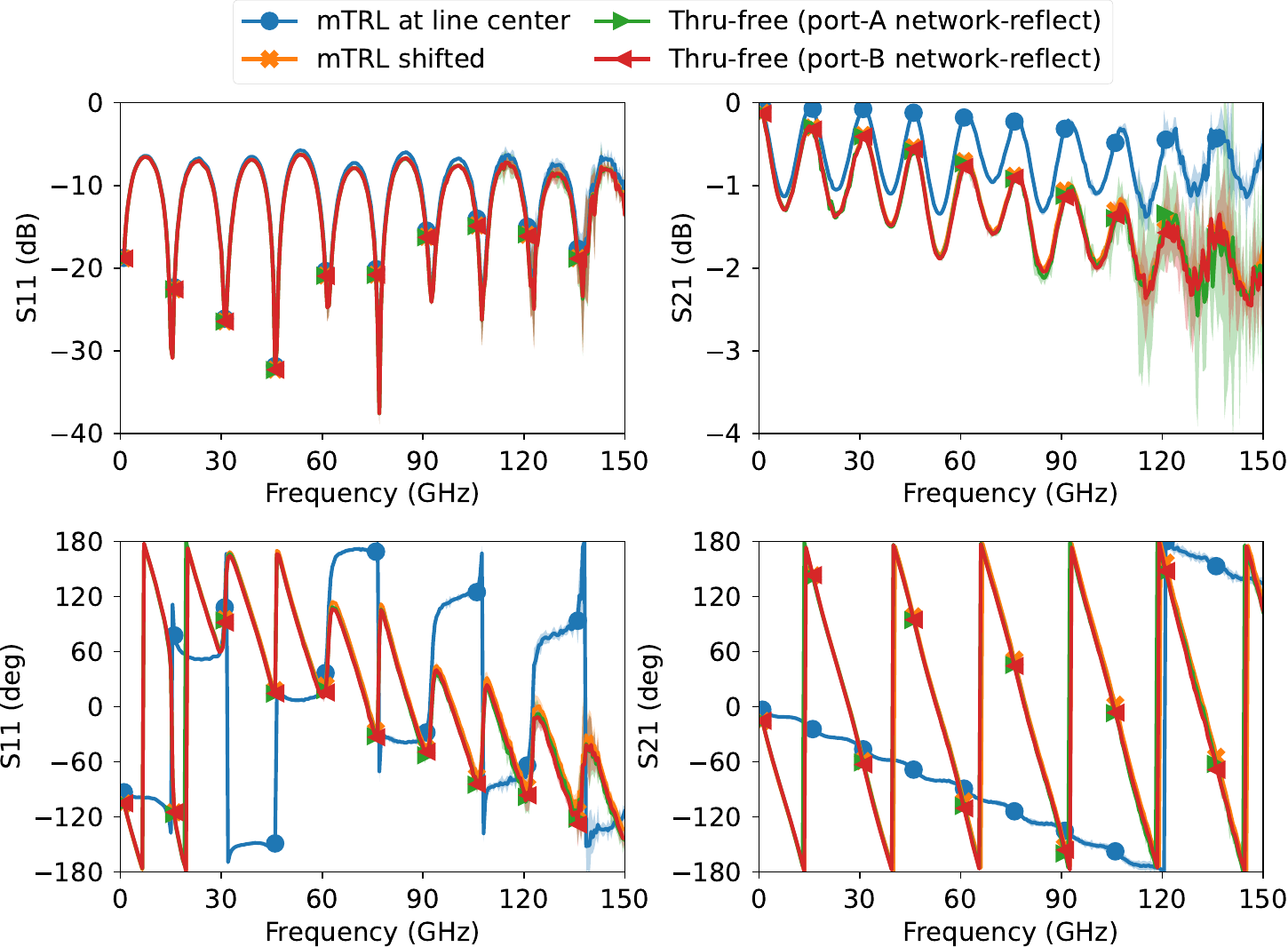}
	\caption{Calibrated measurement of a $6\,\mathrm{mm}$ long $30\,\Omega$ stepped-impedance line. The reference line used in multline TRL calibration has a length of $6.5\,\mathrm{mm}$. The uncertainty bounds correspond to a 95\,\% coverage of a Gaussian distribution due to noise from the VNA propagated linearly through the calibrations.}
	\label{fig:4.4}
\end{figure}

As an additional analysis, we selected  a line with a non-zero length as the reference in the multiline TRL calibration. In the example mentioned previously, the reference line was a thru standard. Thus, post-processing to shift the calibration plane was unnecessary. For the current example, we choose the $6.5\,\mathrm{mm}$ line as the reference line in multiline TRL calibration. The calibrated DUT result is shown in Fig.~\ref{fig:4.4}. As the plot shows, we need to shift the calibration plane backward using the propagation constant derived from the calibration to establish the reference plane at the desired location. However, for the thru-free method, no changes are made, and the calibration plane is automatically set by the measured network, network-reflect, and reflect standards. Therefore, in the thru-free method, we establish the calibration plane location using physical artifacts, whereas in multiline TRL, if a thru standard is not utilized, we must shift the calibration plane location in post-processing utilizing the derived propagation constant.


%% file: Sections/Section5.tex
\section{Conclusion}
\label{sec:5}

We presented a modified version of multiline TRL calibration that eliminates the need for explicitly defining a thru standard. The proposed thru-free multiline calibration was compared to multiline TRL using measurements of microstrip lines fabricated on a PCB with a stepped-impedance DUT for verification. We observed excellent agreement between the proposed method and the multiline TRL calibration when a thru standard was used to set the reference plane.

In cases where a thru standard is not available, the multiline TRL method requires shifting the calibration plane in post-processing to the desired location. This is in contrast to the proposed method, where the location of the calibration plane is set automatically by the measured artifacts. The advantage of the proposed thru-free method is that it eliminates the requirement to explicitly define a thru standard in multiline TRL calibration, making all calibration standards in the thru-free method partially defined.

%% file: Sections/AppendixA.tex
\section{Port Uncertainty of Anritsu ME7838D VNA}
\label{anx:A}

\begin{figure}[th!]
	\centering
	\includegraphics[width=0.95\linewidth]{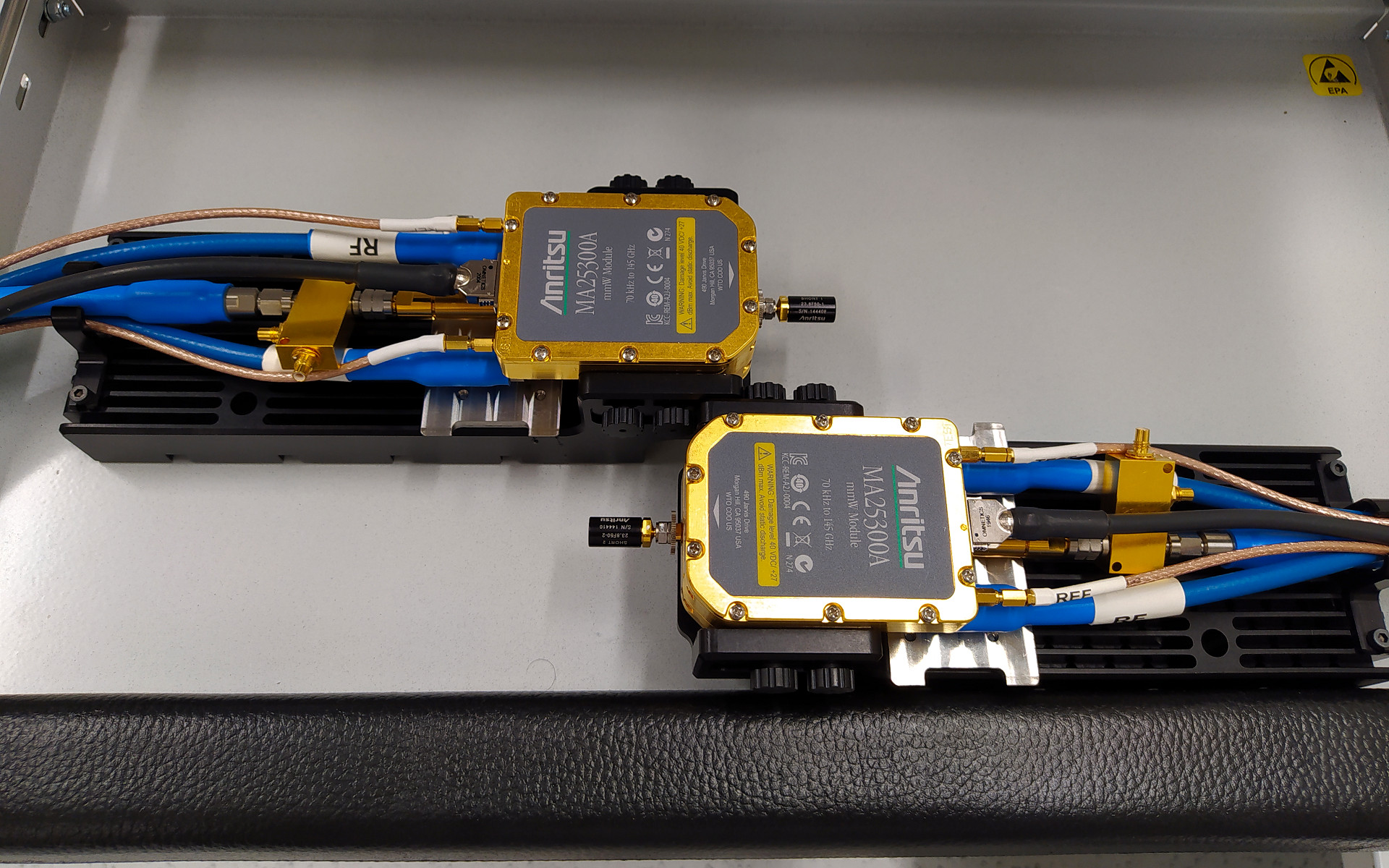}
	\caption{Measurement setup depicting the mm-wave extenders with coaxial 0.8\,mm short standards connected to them.}
	\label{fig:A.1}
\end{figure}

The purpose of this section is to draw attention to the imbalance in noise uncertainty between the two ports of the Anritsu ME7838D VNA used for the measurements discussed in this paper. The test measurement for evaluating the uncertainty of each port was fairly straightforward. We connected a 0.8\,mm coaxial short standard to each port, as shown in Fig.~\ref{fig:A.1}. The short standard was measured while the VNA was in an uncalibrated state. The measurement was performed in four configurations, with power levels -10\,dBm and -20\,dBm, and IF-bandwidths of 100\,Hz and 1\,kHz. To evaluate the statistics of the VNA, a frequency sweep between 1\,GHz and 150\,GHz was conducted, 100 times for the 100\,Hz IF-bandwidth and 500 times for the 1\,kHz IF-bandwidth. 

\begin{figure}[th!]
	\centering
	\includegraphics[width=1\linewidth]{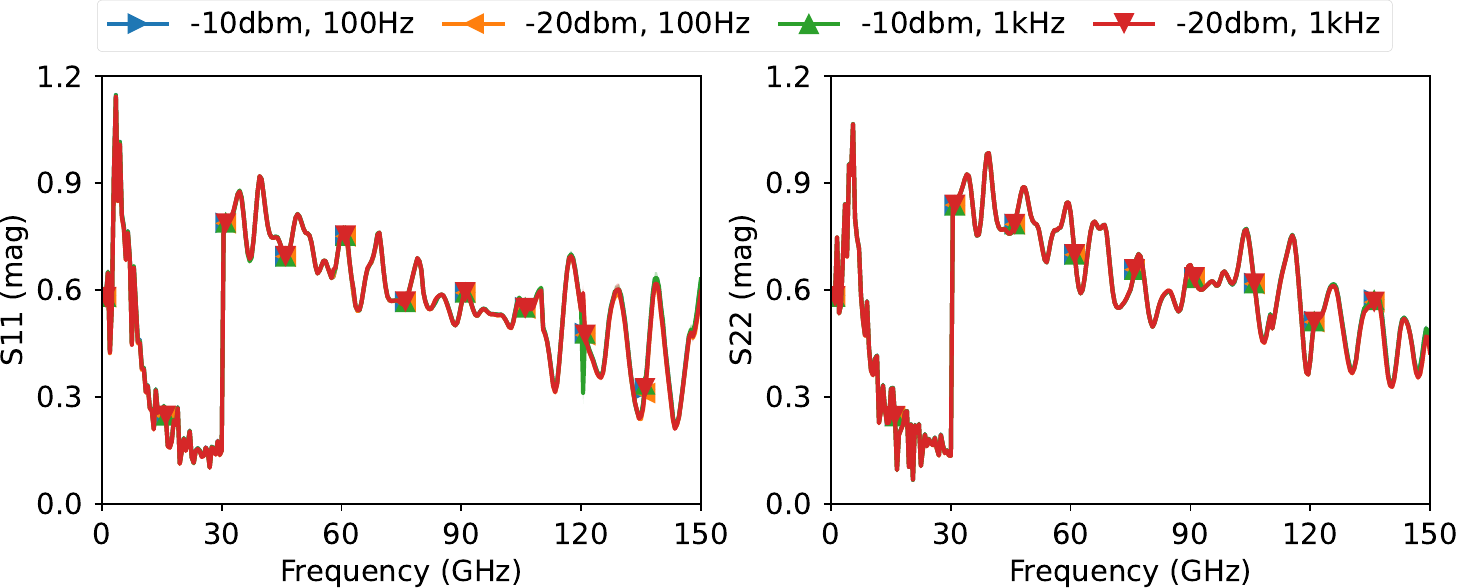}
	\caption{Mean-value of the raw measurement of the 0.8\,mm coaxial short standard under different VNA configurations.}
	\label{fig:A.2}
\end{figure}

\begin{figure}[th!]
	\centering
	\includegraphics[width=1\linewidth]{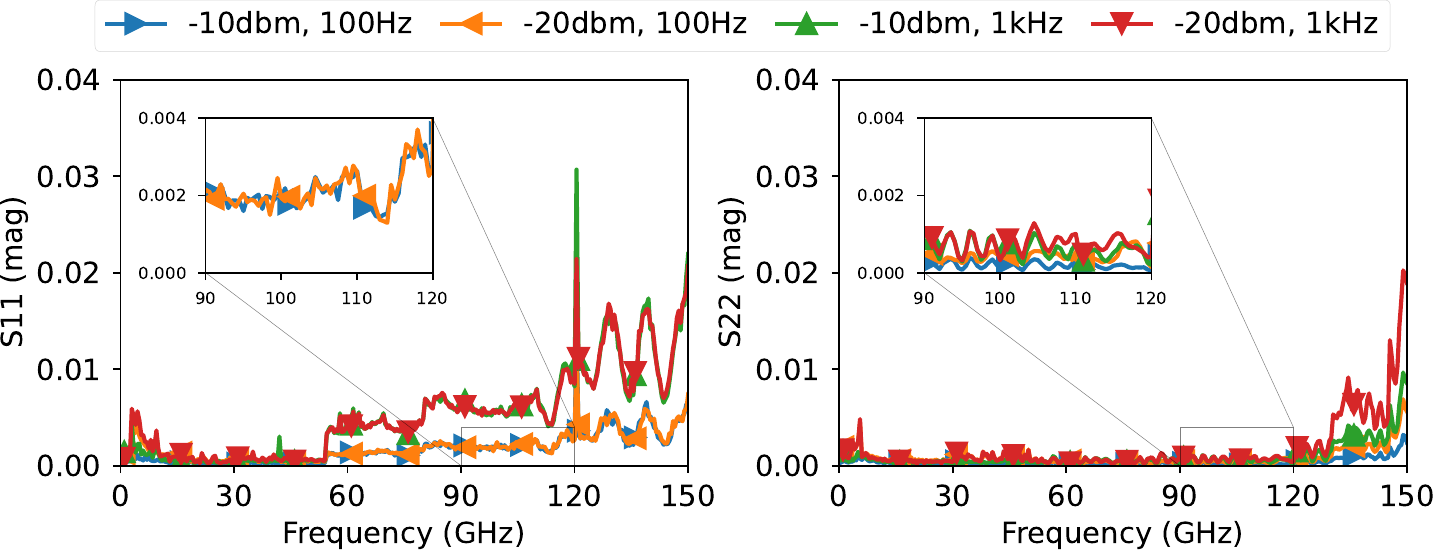}
	\caption{Uncertainty of the raw measurement of the 0.8\,mm coaxial short standard. The uncertainty is reported as the 95\,\% coverage of a Gaussian distribution.}
	\label{fig:A.3}
\end{figure}

In Fig.~\ref{fig:A.2}, we present the mean value of the measured short standard. Across all configurations, there appears to be no difference between the ports. However, in Fig.~\ref{fig:A.3}, we show the standard deviation of the measurements, which clearly indicates a significant noise contribution in port 1 (port $\bs{A}$), in comparison to port 2 (port $\bs{B}$).

The uncertainty jump in the $S_{11}$ measurement start at $54\,\mathrm{GHz}$, which is where the power level settings of the Anritsu ME7838D VNA split. This VNA has two power level settings, one for frequencies below $54\,\mathrm{GHz}$ and the other for frequencies above this value. Although Fig.~\ref{fig:A.3} already demonstrates the poor statistical performance of port 1 compared to port 2, we can see a clear difference in the uncertainty of the traces at 110\,GHz when the settings are -10\,dBm and 100\,Hz. Specifically, port 1 yields an expanded uncertainty of 0.00132, while port 2 yields an expanded uncertainty of 0.00011, which is a factor of 10 difference between the two ports. This difference scales even further during calibration, as demonstrated in the measurements presented in Section \ref{sec:4}.

%% file: Sections/AppendixB.tex
\section{Deriving Calibrated S-parameters}
\label{anx:B}

The calibrated S-parameters can be computed easily by multiplying the inverse of the error boxes as T-parameters and then converting them to S-parameters. This can be expressed as follows:
\begin{equation}
	\bs{S}_\mathrm{cal} = \ttos{\frac{1}{k}\bs{A}^{-1}\stot{\bs{S}_\mathrm{raw}}\bs{B}^{-1}},
	\label{eq:B.1}
\end{equation}
where $\bs{S}_\mathrm{cal}$ and $\bs{S}_\mathrm{raw}$ represent the calibrated and raw measurements of the S-parameter of an arbitrary DUT.

Applying the equation above, the calibrated S-parameters can be expressed as follows:
\begin{subequations}
	\begin{align}
		S_{11}^\mathrm{cal} &= \frac{b_{12} \left(\det\left(\bs{S}_\mathrm{raw}\right) - a_{12} S^\mathrm{raw}_{22}\right) - b_{11} \left(a_{12} - S^\mathrm{raw}_{11}\right) }{b_{11} \left(a_{11} - a_{21} S^\mathrm{raw}_{11}\right) + b_{12} \left(a_{11}S^\mathrm{raw}_{22} - \det\left(\bs{S}_\mathrm{raw}\right) a_{21}\right)},\raisetag{-1ex}\label{eq:B.2a}\\[5pt]
		S_{21}^\mathrm{cal} &= \frac{k S^\mathrm{raw}_{21} \left(a_{11} - a_{12} a_{21}\right) \left(b_{11} - b_{12} b_{21}\right)}{b_{11} \left(a_{11} - a_{21} S^\mathrm{raw}_{11}\right) + b_{12} \left(a_{11}S^\mathrm{raw}_{22} - \det\left(\bs{S}_\mathrm{raw}\right) a_{21}\right)},\raisetag{-1ex}\label{eq:B.2b}\\[5pt]
		S_{12}^\mathrm{cal} &= \frac{S^\mathrm{raw}_{12}/k}{b_{11} \left(a_{11} - a_{21} S^\mathrm{raw}_{11}\right) + b_{12} \left(a_{11}S^\mathrm{raw}_{22} - \det\left(\bs{S}_\mathrm{raw}\right) a_{21}\right)},\raisetag{-1ex}\label{eq:B.2c}\\[5pt]
		S_{22}^\mathrm{cal} &= \frac{a_{11} \left(b_{21} + S^\mathrm{raw}_{22}\right) - a_{21} \left(\det\left(\bs{S}_\mathrm{raw}\right) + b_{21} S^\mathrm{raw}_{11}\right)}{b_{11} \left(a_{11} - a_{21} S^\mathrm{raw}_{11}\right) + b_{12} \left(a_{11}S^\mathrm{raw}_{22} - \det\left(\bs{S}_\mathrm{raw}\right) a_{21}\right)},\raisetag{-1ex}\label{eq:B.2d}
	\end{align}
	\label{eq:B.2}
\end{subequations}
where $\det\left(\bs{S}_\mathrm{raw}\right) = S^\mathrm{raw}_{11} S^\mathrm{raw}_{22} - S^\mathrm{raw}_{12} S^\mathrm{raw}_{21}$.

The expressions for calibrated $S_{21}$ and $S_{12}$ do indeed show dependence on $a_{11}$ and $b_{11}$. However, simplifying the expressions reveals that the calibrated $S_{21}$ and $S_{12}$ only depend on the normalized error terms obtained from the eigenvalue formulation $\{a_{12}, a_{21}/a_{11}, b_{21}, b_{12}/b_{11}\}$, the combined error term $a_{11}b_{11}$, and the transmission error term $k$, which are obtained from the thru measurement, as given by \eqref{eq:2.4}. The expressions for $S^\mathrm{cal}_{21}$ and $S^\mathrm{cal}_{12}$ can be rewritten and simplified as follows:
\begin{equation}
	S^\mathrm{cal}_{21} = \frac{kS_{21}^\mathrm{raw}u}{v}, \qquad S^\mathrm{cal}_{12} = \frac{S_{12}^\mathrm{raw}/k}{v},
	\label{eq:B.4}
\end{equation}
where the numerator $u$ and denominator $v$ are given by
\begin{subequations}
	\begin{align}
		u &= a_{11}b_{11}\left(1 - a_{12} \frac{a_{21}}{a_{11}}\right) \left(1 - \frac{b_{12}}{b_{11}} b_{21}\right), \label{eq:B.5a}\\[5pt]
		v &= a_{11}b_{11}\left[1-\frac{a_{21}}{a_{11}}S_{11}^\mathrm{raw} + \frac{b_{12}}{b_{11}}\left(S_{22}^\mathrm{raw}-\det\left(\bs{S}_\mathrm{raw}\right)\frac{a_{21}}{a_{11}}\right) \right].\raisetag{-1ex}\label{eq:B.5b}
	\end{align}
	\label{eq:B.5}
\end{subequations}

The expressions for $u$ and $v$ show that $S^\mathrm{cal}_{21}$ and $S^\mathrm{cal}_{12}$ indeed depend solely on the normalized error terms $\{a_{12}, a_{21}/a_{11}, b_{21}, b_{12}/b_{11}\}$, the combined error term $a_{11}b_{11}$, and the transmission error term $k$. This means that the terms $a_{11}$ and $b_{11}$ are never found separately, which explains why the uncertainty due to the reflect standard is nullified in multiline TRL calibration as observed in Fig.~\ref{fig:4.3}.